\documentstyle[12pt,latexsym,amssymb,graphicx,axodraw,epsfig,epsf,feynmp,rotating]{article}

\catcode`@=11
\def\citer{\@ifnextchar
[{\@tempswatrue\@citexr}{\@tempswafalse\@citexr[]}}

\def\@citexr[#1]#2{\if@filesw\immediate\write\@auxout{\string\citation{#2}}\fi
  \def\@citea{}\@cite{\@for\@citeb:=#2\do
    {\@citea\def\@citea{--\penalty\@m}\@ifundefined
       {b@\@citeb}{{\bf ?}\@warning
       {Citation `\@citeb' on page \thepage \space undefined}}%
\hbox{\csname b@\@citeb\endcsname}}}{#1}}
\catcode`@=12

\newcommand{\non}{\nonumber}

\newcommand{\beq}{\begin{eqnarray}}
\newcommand{\eeq}{\end{eqnarray}}
\newcommand{\bq}{\begin{equation}}
\newcommand{\eq}{\end{equation}}
\newcommand{\be}{\begin{equation}}
\newcommand{\ee}{\end{equation}}

\begin{document}

\newcommand{\br}{\begin{eqnarray}}
\newcommand{\er}{\end{eqnarray}}
\newcommand{\ba}{\begin{array}}
\newcommand{\ea}{\end{array}}
\newcommand{\bi}{\begin{itemize}}
\newcommand{\ei}{\end{itemize}}
\newcommand{\bn}{\begin{enumerate}}
\newcommand{\en}{\end{enumerate}}
\newcommand{\bc}{\begin{center}}
\newcommand{\ec}{\end{center}}
\newcommand{\ul}{\underline}
\newcommand{\ol}{\overline}
\def\epem{\ifmmode{e^+ e^-} \else{$e^+ e^-$} \fi}
\newcommand{\eeww}{$e^+e^-\rightarrow W^+ W^-$}
\newcommand{\qqQQ}{$q_1\bar q_2 Q_3\bar Q_4$}
\newcommand{\eeqqQQ}{$e^+e^-\rightarrow q_1\bar q_2 Q_3\bar Q_4$}
\newcommand{\eewwqqqq}{$e^+e^-\rightarrow W^+ W^-\ar q\bar q Q\bar Q$}
\newcommand{\eeqqgg}{$e^+e^-\rightarrow q\bar q gg$}
\newcommand{\eeqloop}{$e^+e^-\rightarrow q\bar q gg$ via quark loops}
\newcommand{\eeqqqq}{$e^+e^-\rightarrow q\bar q Q\bar Q$}
\newcommand{\eewwjjjj}{$e^+e^-\rightarrow W^+ W^-\rightarrow 4~{\rm{jet}}$}
\newcommand{\eeqqggjjjj}{$e^+e^-\rightarrow q\bar 
q gg\rightarrow 4~{\rm{jet}}$}
\newcommand{\eeqloopjjjj}{$e^+e^-\rightarrow q\bar 
q gg\rightarrow 4~{\rm{jet}}$ via quark loops}
\newcommand{\eeqqqqjjjj}{$e^+e^-\rightarrow q\bar q Q\bar Q\rightarrow
4~{\rm{jet}}$}
\newcommand{\eejjjj}{$e^+e^-\rightarrow 4~{\rm{jet}}$}
\newcommand{\jjjj}{$4~{\rm{jet}}$}
\newcommand{\qqbar}{$q\bar q$}
\newcommand{\ww}{$W^+W^-$}
\newcommand{\ar}{\rightarrow}
\newcommand{\sm}{${\cal {SM}}$}
\newcommand{\Dir}{\kern -6.4pt\Big{/}}
\newcommand{\Dirin}{\kern -10.4pt\Big{/}\kern 4.4pt}
\newcommand{\DDir}{\kern -7.6pt\Big{/}}
\newcommand{\DGir}{\kern -6.0pt\Big{/}}
\newcommand{\wwqqqq}{$W^+ W^-\ar q\bar q Q\bar Q$}
\newcommand{\qqgg}{$q\bar q gg$}
\newcommand{\qloop}{$q\bar q gg$ via quark loops}
\newcommand{\qqqq}{$q\bar q Q\bar Q$}
\newcommand{\ord}{{\cal O}}
\newcommand{\Ecm}{E_{\mathrm{cm}}}

\def\l{\left\langle}
\def\r{\right\rangle}
\def\aem{\alpha_{\rm em}}
\def\as{\alpha_{\rm s}}
\def\MW{m_{W^\pm}}
\def\MZ{m_{Z}}
\def\ycut{y_{\rm cut}}
\def\Ord{\lower .7ex\hbox{$\;\stackrel{\textstyle <}{\sim}\;$}}
\def\OOrd{\lower .7ex\hbox{$\;\stackrel{\textstyle >}{\sim}\;$}}

\begin{flushright}
DESY 99-192\\
RAL-TR-99-083\\
February 2000
\end{flushright}
\vskip1.0cm
\begin{center}{\Large\bf
Double Higgs production at TeV Colliders\\[0.35cm]
in the Minimal Supersymmetric Standard Model}
\end{center}
\vskip0.75cm
\begin{center}{\large
R. Lafaye$^1$, 
D.J. Miller$^2$, M. M\"uhlleitner$^2$ and  S.~Moretti$^3$}
\end{center}
\vskip0.5cm
\begin{center}{
{\it 1) Laboratoire d'Annecy-le-Vieux de Physique des Particules (LAPP)},\\
{\it B.P. 110, F-74941 Annecy-le-Vieux CEDEX, France.}\\[0.35cm]
{\it 2) Deutsches Elektronen-Synchrotron (DESY),}\\
{\it Notkestrasse 85, D-22603 Hamburg, Germany.}\\[0.35cm]
{\it 3) Rutherford Appleton Laboratory (RAL),}\\
{\it Chilton, Didcot, Oxon OX11 0QX, United Kingdom.}\\[1.0cm]
{\sl Contribution to the Workshop `Physics at TeV Colliders',}\\
{\sl Les Houches, France, 8--18 June 1999}\\
{\sl (to appear in the proceedings)}\\
}
\end{center}

%
%

\vskip1.0cm
\begin{abstract}
  The reconstruction of the Higgs potential in the Minimal
  Supersymmetric Standard Model (MSSM) requires the measurement of the
  trilinear Higgs self-couplings.  The `double Higgs production'
  subgroup has been investigating the possibility of detecting
  signatures of processes carrying a dependence on these vertices at
  the Large Hadron Collider (LHC) and future Linear Colliders (LCs).
  As reference reactions, we have chosen $gg\to hh$ and $e^+e^-\to h h
  Z$, respectively, where $h$ is the lightest of the MSSM Higgs
  bosons.  In both cases, the $Hhh$ interaction is involved.  For
  $m_H\OOrd2m_h$, the two reactions are resonant in the $H\to hh$
  mode, providing cross sections which are detectable at both
  accelerators and strongly sensitive to the strength of the trilinear
  coupling involved. We explore this mass regime of the MSSM in the
  $h\to b\bar b$ decay channel, also accounting for irreducible
  background effects.
\end{abstract}
\newpage

\section{Introduction}
\label{sec_intro}

Considerable attention has been devoted to double Higgs boson
production at future $e^+e^-$ and hadron colliders, both in the
Standard Model (SM) and the MSSM (see Refs.~\cite{revsmee,revsmlhc} for
an incomplete list of references for SM $e^+e^-$ and hadron colliders,
respectively, and similarly \cite{revmssmee,revmssmlhc} for the MSSM).  
For the SM,
detailed signal-to-background studies already exist for a LC
environment \cite{ee}, for both `reducible' and `irreducible'
backgrounds \cite{Lutz,noi}, which have assessed the feasibility of
experimental analyses.  At the LHC, since here the typical SM signal
cross sections are of the order of 10 fb \cite{pp}, high integrated
luminosities would be needed to generate a statistically large enough
sample of double Higgs events. These would be further obscured by an
overwhelming background, making their selection and analysis in a
hadronic environment extremely difficult.  Thus, in this contribution we will
concentrate only on the case of the MSSM.

In the Supersymmetric (SUSY) scenario, the phenomenological potential
of these reactions is two-fold. Firstly, in some specific cases, they
can furnish new discovery channels for Higgs bosons.  Secondly, they
are all dependent upon several triple Higgs self-couplings of the theory,
which can then be tested by comparing theoretical predictions with
experimental measurements. This is the first step in the
reconstruction of the Higgs potential itself\footnote{The
  determination of the quartic self-interactions is also required, but
  appears out of reach for some time: see Refs.~\cite{ee,pp} for some
  cross sections of triple Higgs production.}.

The Higgs Working Group (WG) has focused much of its attention in assessing
the viability of these reactions at future TeV colliders. However, the
number of such processes is very large both at the LHC and a LC
\cite{ee,pp}, so only a few `reference' reactions could be studied in
the context of this Workshop.  Work is in progress for the
longer term, which aims to cover most of the double Higgs production
and decay phenomenology at both accelerators~\cite{more}.

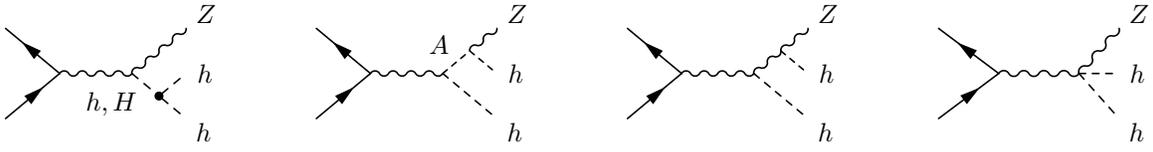
\begin{figure}[!t]
\begin{flushleft}
\underline{$gg$ to double Higgs fusion at the LHC: $gg\to hh$}
\\[1.5\baselineskip]
{\footnotesize
\unitlength1mm
\hspace{5mm}
\begin{fmfshrink}{0.7}
\begin{fmfgraph*}(30,12)
  \fmfstraight
  \fmfleftn{i}{2} \fmfrightn{o}{2}
  \fmflabel{$g$}{i1}  \fmflabel{$g$}{i2}
  \fmf{gluon,tens=2/3}{i1,v1} \fmf{phantom}{v1,v2,v3,o1}
  \fmf{gluon,tens=2/3}{w1,i2} \fmf{phantom}{w1,w2,w3,o2}
  \fmffreeze
  \fmf{fermion}{w1,x2,v1}
  \fmf{dashes, lab=$h,,H$}{x2,x3}
  \fmf{dashes}{o1,x3,o2}
  \fmffreeze
  \fmf{fermion,label=$t,,b$,label.s=left}{v1,w1}
  \fmflabel{$h$}{o1}  \fmflabel{$h$}{o2}
  \fmfdot{x3}
\end{fmfgraph*}
\hspace{15mm}
\begin{fmfgraph*}(30,12)
  \fmfstraight
  \fmfleftn{i}{2} \fmfrightn{o}{2}
  \fmf{gluon}{i1,v1} \fmf{phantom}{v1,v3} \fmf{dashes}{v3,o1}
  \fmf{gluon}{w1,i2} \fmf{phantom}{w1,w3} \fmf{dashes}{w3,o2}
  \fmffreeze
  \fmf{fermion}{v1,w1,w3,v3,v1}
  \fmflabel{$h$}{o1}  \fmflabel{$h$}{o2}
\end{fmfgraph*}
\end{fmfshrink}
}
\\[2\baselineskip]
\underline{double Higgs-strahlung at a LC: $e^+e^-\to hhZ$}
\\[1.5\baselineskip]
{\footnotesize
\unitlength1mm
\hspace{5mm}
\begin{fmfshrink}{0.7}
\begin{fmfgraph*}(24,12)
  \fmfstraight
  \fmfleftn{i}{3} \fmfrightn{o}{3}
  \fmf{fermion}{i1,v1,i3}
  \fmf{boson,tens=3/2}{v1,v2}
  \fmf{boson}{v2,o3} \fmflabel{$Z$}{o3}
  \fmf{phantom}{v2,o1}
  \fmffreeze
  \fmf{dashes,lab=$h,,H$,lab.s=right}{v2,v3} \fmf{dashes}{v3,o1}
  \fmffreeze
  \fmf{dashes}{v3,o2} 
  \fmflabel{$h$}{o2} \fmflabel{$h$}{o1}
  \fmfdot{v3}
\end{fmfgraph*}
\hspace{15mm}
\begin{fmfgraph*}(24,12)
  \fmfstraight
  \fmfleftn{i}{3} \fmfrightn{o}{3}
  \fmf{fermion}{i1,v1,i3}
  \fmf{boson,tens=3/2}{v1,v2}
  \fmf{dashes}{v2,o1} \fmflabel{$H$}{o1}
  \fmf{phantom}{v2,o3}
  \fmffreeze
  \fmf{dashes,lab=$A$,lab.s=left}{v2,v3} 
  \fmf{boson}{v3,o3} \fmflabel{$Z$}{o3}
  \fmffreeze
  \fmf{dashes}{v3,o2} 
  \fmflabel{$h$}{o2} \fmflabel{$h$}{o1}
\end{fmfgraph*}
\hspace{15mm}
\begin{fmfgraph*}(24,12)
  \fmfstraight
  \fmfleftn{i}{3} \fmfrightn{o}{3}
  \fmf{fermion}{i1,v1,i3}
  \fmf{boson,tens=3/2}{v1,v2}
  \fmf{dashes}{v2,o1} \fmflabel{$H$}{o1}
  \fmf{phantom}{v2,o3}
  \fmffreeze
  \fmf{boson}{v2,v3,o3} \fmflabel{$Z$}{o3}
  \fmffreeze
  \fmf{dashes}{v3,o2} 
  \fmflabel{$h$}{o2} \fmflabel{$h$}{o1}
\end{fmfgraph*}
\hspace{15mm}
\begin{fmfgraph*}(24,12)
  \fmfstraight
  \fmfleftn{i}{3} \fmfrightn{o}{3}
  \fmf{fermion}{i1,v1,i3}
  \fmf{boson,tens=3/2}{v1,v2}
  \fmf{dashes}{v2,o1} \fmflabel{$h$}{o1}
  \fmf{dashes}{v2,o2} \fmflabel{$h$}{o2}
  \fmf{boson}{v2,o3} \fmflabel{$Z$}{o3}
\end{fmfgraph*}
\end{fmfshrink}
}
\end{flushleft}
\caption{
Diagrams contributing to $gg\to hh$ (top) and $e^+e^-\to hhZ$ 
(bottom) in the MSSM. 
}
\label{fig:graphs}
\end{figure}

These reference reactions were chosen to be $gg\to hh$ for the LHC
(see top of Fig.~\ref{fig:graphs}) and $e^+e^-\to h h Z$ for the LC
(see bottom of Fig.~\ref{fig:graphs}), where $h$ is the lightest of the
MSSM scalar Higgs bosons.  The reason for this preference is simple.
Firstly, a stable upper limit exists on the value of $m_h$, of the
order of 130 GeV, now at two-loop level \cite{twoloop}, so that
its detection is potentially well within the reach of both the LHC and
a LC.  In contrast, the mass of all other Higgs bosons of the MSSM may
vary from the electroweak (EW) scale, ${\cal O}(m_Z)$, up to the TeV
region.  In addition, as noted in Ref.~\cite{pp}, the multi-$b$ final
state in $gg\to hh\to b\bar b b\bar b$, with two resonances and large
transverse momenta, may be exploited in the search for the $h$ scalar
in the large $\tan\beta$ and moderate $m_A$ region. This is a corner
of the MSSM parameter space that has so far eluded the scope of the
standard Higgs production and decay modes \cite{standard}.  (The
symbol $A$ here denotes the pseudoscalar Higgs boson of the MSSM, and
we reserve the notation $H$ for the heaviest scalar Higgs state of the
model.)
However, this paper will not investigate the LHC discovery potential
in this mode, given the very sophisticated treatment of the background
(well beyond the scope of this note) required by the assumption
that no $h$ scalar state has been previously discovered (see below).
This will be done in Ref.~\cite{more}.  Furthermore, the $gg\to hh$
and $e^+e^-\to hhZ$ modes largely dominate double Higgs production
\cite{ee,pp}, at least for centre-of-mass (CM) energies of 14 TeV
at the LHC and 500 GeV in the case of a LC, the default values of
our analysis. (Notice that we assume no
polarization of the incoming beams in $e^+e^-$ scatterings.)
 Finally, when $m_H\OOrd2m_h$, the two reactions are
resonant, as they can both proceed via intermediate states involving
$H$ scalars, through $gg\to H$ and $e^+e^-\to HZ$, which in turn decay
via $H\to hh$ \cite{BRs}. Thus, the production cross sections are
largely enhanced \cite{ee,pp} (up to two orders of magnitude above
typical SM rates at the LHC \cite{pp}) and become clearly visible. 
This allows the possibility of probing the
trilinear $Hhh$ vertex at one or both these colliders.

The dominant decay rate of the MSSM $h$ scalar is into $b\bar b$
pairs, regardless of the value of $\tan\beta$ \cite{BRs}.  Therefore,
the final signatures of our reference reactions always involve four
$b$-quarks in the final state. (In the case of a LC environment, a
further trigger on the accompanying $Z$ boson can be exploited.)  

If one assumes very efficient tagging and high-purity sampling of
$b$-quarks, the background to $hh$ events at the LHC is dominated by
the irreducible QCD modes \cite{ATLTDR}. Among these, we focus here 
on the cases $q\bar q,gg\to b\bar b b\bar b$, as representative of
ideal $b$-tagging performances. These modes
consist of a purely QCD contribution
of ${\cal O}(\alpha_s^4)$, an entirely EW process of ${\cal
  O}(\alpha_{em}^4)$ (with no double
Higgs intermediate states) and an ${\cal O}(\alpha_s^2\alpha_{em}^2)$
component consisting of EW and QCD interactions.  (Note that in the EW
case only $q\bar q$ initiated subprocesses are allowed at tree-level.)
For a LC, the final state of the signal  is $ b\bar b b\bar b Z$, with
the $Z$ reconstructed from its decay products in some channel. Here, the EW
background is of $\ord(\alpha_{em}^5)$ away from resonances
(and, again, contains no more than one intermediate Higgs boson), whereas
the EW/QCD background is proportional to $(\alpha_s^2\alpha_{em}^3)$.

In general, EW backgrounds can be problematic due to the presence of
$Z$ vectors and single Higgs scalars yielding $b \bar b$ pairs, with
the partons being typically at large transverse momenta and well separated.  In
contrast, the QCD backgrounds involve no heavy objects decaying to $b
\bar b$ pairs and are dominated by the typical infrared (i.e., soft
and collinear) QCD behaviour of the partons in the final state.
However, they can yield large production rates because of the strong
couplings.
\begin{fmffile}{fd}
\end{fmffile}

In this study, we investigate the interplay between the signal and
background at both colliders, adopting detector as well as dedicated
selection cuts. We carry out our analysis at both parton and hadron
level.  The plan of this note is as follows. The next Section details
the procedure adopted in the numerical computation.
Sect.~\ref{sec_results} displays our results and contains our
discussion. Finally, in the last section, we summarise our findings
and consider possible future studies.

\section{Calculation}
\label{sec_calculation}

For the parton level simulation, the double Higgs production process
at the LHC, via $gg$ fusion, has been simulated using the program
of Ref.~\cite{spira} to generate the interaction $gg \to hh$,
with the matrix elements (MEs) taken at leading-order (LO) for consistency
with our treatment of the background. We then
perform the two $h\to b\bar b$ decays to obtain the actual $4b$-final
state.  For double Higgs production at a LC, we use a source
code for the signal derived from that already used in Ref.~\cite{noi}. 
At both colliders,
amplitudes for background events were generated by means of MadGraph
\cite{tim} and the {\tt HELAS} package \cite{HELAS}. Note that
interferences between signal and backgrounds, and between the various
background contributions themselves, have been neglected.  This is a
good approximation for the interferences involving the signal because
of the very narrow width of the MSSM lightest Higgs boson.  Similarly,
the various background subprocesses have very different topologies,
and one would expect their interferences to be small in general.
 
The Higgs boson masses and couplings of the MSSM can be expressed at
tree-level in terms of the mass of the pseudoscalar Higgs state,
$m_A$, and the ratio of the vacuum expectation values of the two neutral
fields in the two iso-doublets, $\tan\beta$.
At higher order however, top and stop
loop-effects can become significant.  Radiative corrections in the
one-loop leading $m_t^4$ ap\-pro\-xi\-ma\-tion are parameterised by
\beq
\epsilon \approx \frac{3 G_F m_t^4}{\sqrt{2} \pi^2 \sin^2 \beta} 
\log \frac{m_S^2}{m_t^2} 
\eeq
where the SUSY breaking scale is given by the common squark mass,
$m_S$, set equal to $1$~TeV in the numerical analysis. If stop mixing
effects are modest at the SUSY scale, they can be accounted for by
shifting $m_S^2$ in $\epsilon$ by the amount $\Delta m_S^2 = \hat{A}^2
[1-\hat{A}^2/(12 m_S^2)]$
($\hat{A}$ is the trilinear common coupling). 
The charged and neutral CP-even Higgs boson
masses, and the Higgs mixing angle $\alpha$ are given in this
approximation by the relations:
\begin{eqnarray}
m_{H^\pm}^2 \!\!&=&\!\!  m_A^2 + 
m_Z^2 \cos^2 \theta_W, \non\\
m_{h,H}^2 \!\!&=&\!\! \textstyle{\frac{1}{2}}
[ m_A^2+m_Z^2+\epsilon \non\\
&\mp&
\sqrt{(m_A^2+m_Z^2+\epsilon)^2- 4m_A^2 m_Z^2 \cos^2 2\beta
   - 4\epsilon( m_A^2 \sin^2 \beta + m_Z^2 \cos^2 \beta)} ],
\non \\
\tan 2\alpha \!\!&=&\!\! \tan 2\beta
 \frac{m_A^2 + m_Z^2}{m_A^2 - m_Z^2 +\epsilon/\cos 2\beta} \qquad
\mbox{with} \qquad  - \frac{\pi}{2} \leq \alpha \leq 0,
\label{mass}
\end{eqnarray}
as a function of $m_A$ and $\tan\beta$.  The triple Higgs self-couplings of
the MSSM can be parameterised \cite{okada,djouadi} in units
of $M_Z^2/v$, $v=246$ GeV, as,
\beq
\lambda_{hhh} &=& 3 \cos2\alpha \sin (\beta+\alpha) 
+ 3 \frac{\epsilon}{m_Z^2} \frac{\cos \alpha}{\sin\beta} \cos^2\alpha,  
\non \\
\lambda_{Hhh} &=& 2\sin2 \alpha \sin (\beta+\alpha) -\cos 2\alpha \cos(\beta
+ \alpha) + 3 \frac{\epsilon}{m_Z^2} \frac{\sin \alpha}{\sin\beta}
\cos^2\alpha, \non \\
\lambda_{HHh} &=& -2 \sin 2\alpha \cos (\beta+\alpha) - \cos 2\alpha \sin(\beta
+ \alpha) + 3 \frac{\epsilon}{m_Z^2} \frac{\cos \alpha}{\sin\beta}
\sin^2\alpha, \non \\
\lambda_{HHH} &=& 3 \cos 2\alpha \cos (\beta+\alpha) 
+ 3 \frac{\epsilon}{m_Z^2} \frac{\sin \alpha}{\sin\beta} \sin^2 \alpha,
\non \\
\lambda_{hAA} &=& \cos 2\beta \sin(\beta+ \alpha)+ 
\frac{\epsilon}{m_Z^2} 
\frac{\cos \alpha}{\sin\beta} \cos^2\beta, \non \\
\lambda_{HAA} &=& - \cos 2\beta \cos(\beta+ \alpha) + 
\frac{\epsilon}{m_Z^2} 
\frac{\sin \alpha}{\sin\beta} \cos^2\beta.
\label{coup}
\eeq

Next-to-leading order (NLO) effects are certainly dominant, though the
next-to-next-to-leading order (NNLO) ones cannot entirely be neglected
(especially in the Higgs mass relations).  Thus, in the numerical
analysis, the complete one-loop and the leading two-loop corrections
to the MSSM Higgs masses and the triple Higgs self-couplings are
included.  The Higgs masses,
widths and self-couplings have been computed using the {\sc HDECAY}
program described in Ref.~\cite{hdecay1}, which uses a running $b$-mass
in evaluating the $h\ar b\bar b$ decay fraction.  Thus, for
consistency, we have evolved the value of $m_b$ entering the $hbb$
Yukawa couplings of the $h\ar b\bar b$ decay currents of our processes
in the same way.

For our analysis, we have considered $\tan \beta=3$ and $50$.  For the
LHC, high values of $\tan \beta$ produce a signal cross section much
larger than the $\tan\beta=3$ scenario, over almost the entire range
of $m_A$. However, this enhancement is due to the increase of the
down-type quark-Higgs coupling, which is proportional to $\tan\beta$
itself, and serves only to magnify the dominance of the quark box
diagrams of Fig.~\ref{fig:graphs}. Unfortunately, these graphs have no
dependence on either of the two triple Higgs self-couplings entering
the gluon-gluon process considered here, i.e., $\lambda_{hhh}$ and
$\lambda_{Hhh}$. Thus, although the cross section is comfortably
observable, all sensitivity to such vertices is lost. Therefore, the
measurement of the triple Higgs self-coupling, $\lambda_{Hhh}$, is only
feasible at the LHC for low $\tan \beta$ due to the resonant
production of the heavy Higgs boson (see Fig. 5a of Ref.~\cite{pp}). 

In contrast, the cross section for double Higgs production at the LC
is small for large $\tan\beta$ because there is no heavy Higgs
resonance (see Fig.~8 of Ref.~\cite{ee}). 
As soon as it becomes kinematically possible to decay the
heavy Higgs into a light Higgs pair, the $ZZH$ coupling is already too
small to generate a sizable cross section. Furthermore, the continuum
MSSM cross section is suppressed with respect to the SM cross section
since the MSSM couplings $ZZH$ and $ZZh$ vary with $\cos(\beta-\alpha)$ and
$\sin(\beta-\alpha)$, respectively, with respect to the corresponding SM
coupling. Notice that in this regime, at a LC, the $\lambda_{hhh}$
vertex could in principle be accessible instead, since
$\lambda_{hhh}\gg \lambda_{Hhh}$ (see Fig.~2 of Ref.~\cite{ee})
and because of the kinematic enhancement induced by $m_h\ll m_H$.
Unfortunately, we will see that the size of the $e^+e^-\to hhZ$ cross
section itself is prohibitively small.

We assume that $b$-jets are distinguishable from light-quark and gluon
jets and no efficiency to tag the four $b$-quarks is included in our
parton level results. We further neglect considering the possibility
of the  $b$-jet charge
determination. Also, to simplify the calculations, the $Z$ boson
appearing in the final state of the LC process is treated as
on-shell and no branching ratio
(BR) is applied to quantify its possible decays. In
practice, one may assume that it decays leptonically (i.e., $Z\to
\ell^+\ell^-$, with $\ell=e,\mu,\tau$) or hadronically into
light-quark jets (i.e., $Z\to q\bar q$, with $q\ne b$), in order to
avoid problems with $6b$-quark combinatorics.  Furthermore, in the LC
analysis, we have not simulated the effects of Initial State Radiation
(ISR), beamstrahlung or Linac energy spread.  Indeed, we expect them
to affect signal and backgrounds rather similarly, so we can neglect
them for the time being. Indeed, since a detailed phenomenological
study, including both hadronisation and detector effects, already
exists for the case of double Higgs-strahlung in $e^+e^-$ \cite{Lutz},
whose conclusions basically support those attained in the
theoretical study of Ref.~\cite{noi}, we limit ourselves here to
update the latter to the case of the MSSM.

So far only resonant production $gg$ $\to$ $H$ $\to$ $hh$ $\to$ 
$b\bar bb\bar b$ has been investigated
\cite{ATLTDR}, with  full hadronic and detector simulation and
considering also the (large) QCD backgrounds, and a similar study
does not exist for continuum double Higgs production at the LHC. 
(See Ref.~\cite{ERW} for a detailed account of the $gg\to H$ $\to$
$hh$ $\to$ $\gamma\gamma b\bar b$ decay channel.)
The event simulation has been
performed by using a special version of {\sc PYTHIA} \cite{pythia1}, in
which the relevant LO MEs for double Higgs production of
Ref.~\cite{spira} have been implemented by M. El Kacimi and R. Lafaye.
These MEs take into account both continuum and resonant
double Higgs boson production and their interferences.
(The insertion of those for $e^+e^-$ processes is in progress.) The
{\sc PYTHIA} interface to {\sc HDECAY} has been exploited in order to
generate the MSSM Higgs mass spectrum and the relevant Higgs BRs, thus
maintaining consistency with the parton level approach.  As for the
LHC detector simulation, the fast simulation package
was used, with high luminosity (i.e., $\int{\cal L}dt=100$ fb$^{-1}$)
parameters. 

The motivation for our study is twofold. On the one
hand, to complement the studies of Ref.~\cite{ATLTDR} by also
considering the continuum production $gg\to hh\to b\bar b b\bar b$
at large $\tan\beta$. On the other hand, to explore the 
possibility of further kinematic suppression
of the various irreducible backgrounds to the resonant channel
at small $\tan\beta$.

\section{Results}
\label{sec_results}

\subsection{The LHC analysis}
\label{subsec_LHC}

In our LHC analysis, following the discussion in Sect.~\ref{sec_calculation},
we focus most of our attention on the case $\tan\beta=3$, with 
$m_A=210$ GeV, although other combinations of these two MSSM parameters
will also be considered.
We further set $A=-\mu=1$ TeV and take all sparticle masses (and other SUSY
scales) to be as large as $1$ TeV.

\subsubsection{$gg\to hh\to b\bar{b}b\bar{b}$ at parton level}
\label{subsubsec_parton_LHC}

In our parton level analysis, we identify jets with the partons from
which they originate (without smearing the momenta) and apply all cuts
directly to the partons.
We mimic the finite coverage of the LHC detectors by imposing a
transverse momentum threshold on each of the four $b$-jets, 
\begin{equation}\label{pTbcut_LHC}
p_T({b}) > 30~{\mathrm{GeV}}
\end{equation}
and requiring their pseudorapidity to be 
\begin{equation}\label{etabcut_LHC}
|\eta({b})| < 2.5.
\end{equation}
Also, to allow for their detection as separate objects, we impose
an isolation criterium among $b$-jets,
\begin{equation}\label{Rbbcut_LHC}
\Delta R({bb})>0.4,
\end{equation}
by means of  the usual cone variable
$\Delta R({ij})=\sqrt{\Delta\eta({ij})^2+\Delta\phi({ij})^2}$,
defined in terms of relative differences in pseudo-rapidity
$\eta_{ij}$ and azimuth $\phi_{ij}$ of the $i$-th and $j$-th $b$-jets.

As preliminary and very basic selection cuts (also to
help the stability of the numerical integration), we have required that
the invariant mass of the entire $4b$-system is at least twice the
mass of the lightest MSSM Higgs boson (apart from mass resolution and
gluon emission effects), e.g., 
\begin{equation}\label{Mbbbbcut_LHC}
m({bbbb})\ge 2m_h-40~{\rm{GeV}},
\end{equation}
and that exactly two $h$-resonances are reconstructed, such that
\begin{equation}\label{Mbbcut_LHC}
|m({bb})-m_h|<20~{\mathrm{GeV}}.
\end{equation}
In doing so, we implicitly assume that the $h$ scalar boson has
already been discovered and its mass measured through some other
channel, as we have already intimated in the Introduction. 

\begin{figure}[!ht]
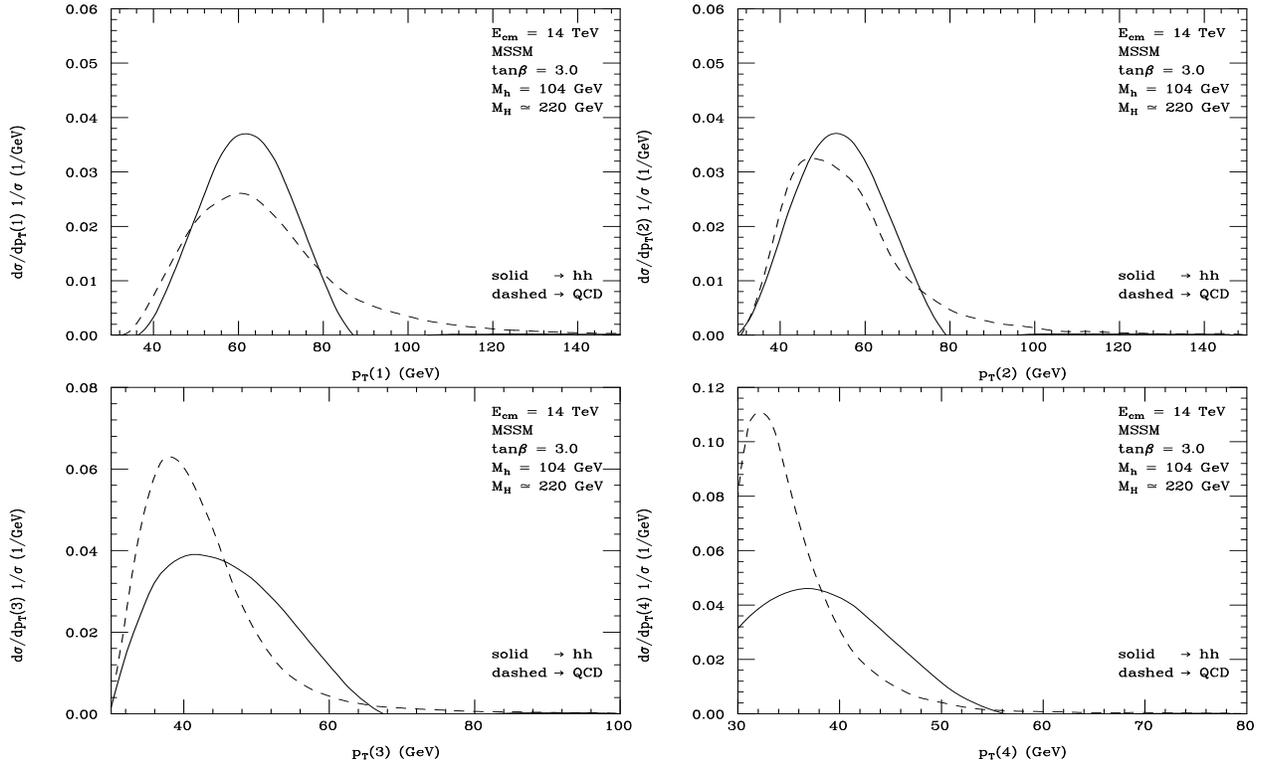

\begin{minipage}[b]{.495\linewidth}
\centering\epsfig{file=pT1_LHC.ps,angle=90,height=5cm,width=\linewidth}
\end{minipage}\hfill\hfill
\begin{minipage}[b]{.495\linewidth}
\centering\epsfig{file=pT2_LHC.ps,angle=90,height=5cm,width=\linewidth}
\end{minipage}\hfill\hfill
\begin{minipage}[b]{.495\linewidth}
\centering\epsfig{file=pT3_LHC.ps,angle=90,height=5cm,width=\linewidth}
\end{minipage}\hfill\hfill
\begin{minipage}[b]{.495\linewidth}
\centering\epsfig{file=pT4_LHC.ps,angle=90,height=5cm,width=\linewidth}
\end{minipage}

\caption{Distributions in transverse momentum of the
four $p_T$-ordered $b$-jets in $gg\to hh\to b\bar b b\bar b$ and
in the QCD background, after the cuts
(\ref{pTbcut_LHC})--(\ref{Mbbcut_LHC}) at the LHC, 
for $\tan\beta=3$, $m_h=104$ GeV and $m_H\simeq220$ GeV.
Normalisation is to unity.}
\vspace*{-3mm}
\label{fig:pTaftercuts_LHC}
\end{figure}

After the above cuts have been implemented, we have found that the two
$4b$-backgrounds proceeding through EW interactions are negligible
compared to the pure QCD process.  In fact, the constraints described
in eqs.~(\ref{Mbbbbcut_LHC})--(\ref{Mbbcut_LHC}) produce the strongest
suppression, almost completely washing out the relatively enhancing
effects that the cuts in (\ref{pTbcut_LHC})--(\ref{Rbbcut_LHC}) have
on the EW components of the backgrounds with respect to the pure QCD
one, owning to the intermediate production of massive $Z$ bosons in
the former. In the end, the production rates of the three subprocesses
scale approximately as their coupling strengths: i.e., ${\cal
  O}(\alpha_s^4)$ : ${\cal O}(\alpha_s^2\alpha_{em}^2)$ : ${\cal
  O}(\alpha_{em}^4)$. Therefore, in the reminder of our analysis, we
will neglect EW effects, as they represent not more than a 10\%
correction to the QCD rates, which are in turn affected by much larger
QCD $K$-factors.  As for the pure QCD background itself, it hugely
overwhelms the double Higgs signal at this stage.  The cross section
of the former is about 7.85 pb, whereas that of the latter is
approximately 0.16 pb.

To appreciate the dominance of the $m_h$ cuts, one may refer to
Fig.~\ref{fig:pTaftercuts_LHC}, where the distributions in transverse
momentum of the four $p_T$-ordered $b$-quarks (such that $p_{T}(b_1)>
... >p_{T}(b_4)$) of both signal and QCD background are shown.  Having
asked the four $b$-jets of the background to closely emulate the
$gg\to hh\to b\bar b b\bar b$ kinematics, it is not surprising to see
a `degeneracy' in the shape of all spectra. Clearly, no
further background suppression can be gained by increasing the
$p_T(b)$ cuts. The same can be said for $\eta({b})$ and $\Delta
R({bb})$. Others quantities ought to be exploited.

\begin{figure}[!ht]
  ~\hskip2.5cm\epsfig{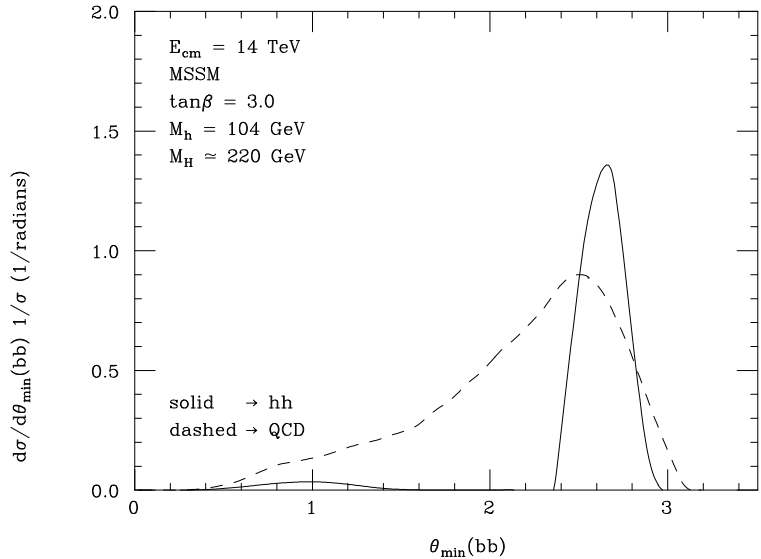}
\caption{Distributions in minimum  relative angle (in radians)
  in the $4b$-system rest frame between two $b$-jets
  reconstructing $m_h$ in $gg\to hh\to b\bar b b\bar b$ and in the QCD
  background, after the cuts (\ref{pTbcut_LHC})--(\ref{Mbbcut_LHC}) at
  the LHC, for $\tan\beta=3$, $m_h=104$ GeV and $m_H\simeq220$ GeV.
  Normalisation is to unity.}
\label{fig:thetabb_LHC}
\vspace*{-3mm}
\end{figure}

In Fig.~\ref{fig:thetabb_LHC}, we present the signal and QCD
background distributions in the minimum angle formed between the
two $b$-quarks coming from the `same Higgs' (i.e., those fulfilling
the cuts in (\ref{Mbbcut_LHC})) in the $4b$-system rest frame (the
plot is rater similar for the maximum angle, thus also on average).
There, one can see a strong tendency of the two $2b$-pairs produced in
the Higgs decays to lie back-to-back, reflecting the $2\to2$
intermediate dynamics of Higgs pair production via $gg\to hh$. Missing
such kinematically constrained virtual state, the QCD background shows
a much larger angular spread towards small $\theta_{\rm{min}}(bb)$
values, eventually tamed by the isolation cut (\ref{Rbbcut_LHC}).

The somewhat peculiar shape of the signal distribution is due to
destructive interference.  Recall that the signal contains not only
diagrams proceeding via a heavy Higgs resonance (the upper-left hand
graph of Fig.~\ref{fig:graphs}), which results in the large peak in
Fig.~\ref{fig:thetabb_LHC}, but also contains a continuum contribution
mediated by box graphs (the upper-right hand graph of
Fig.~\ref{fig:graphs}). These two contributions destructively
interfere leading to the depletion of events between the large
back-to-back peak and the small remaining 'bump' of the continuum
contribution as seen in Fig.~\ref{fig:thetabb_LHC}.

In the end, a good criterium to enhance the signal-to-background ratio
($S/B$) is to require, e.g., $\theta({bb})>2.4$ radians, i.e., a
separation between the $2b$-jets reconstructing the lightest Higgs
boson mass of about 140 degrees in angle. (Incidentally, we also have 
investigated the angle that each of these $2b$-pairs form with the
beam axis, but found no significant difference between signal and QCD
background).

\begin{figure}[!ht]
~\hskip2.5cm\epsfig{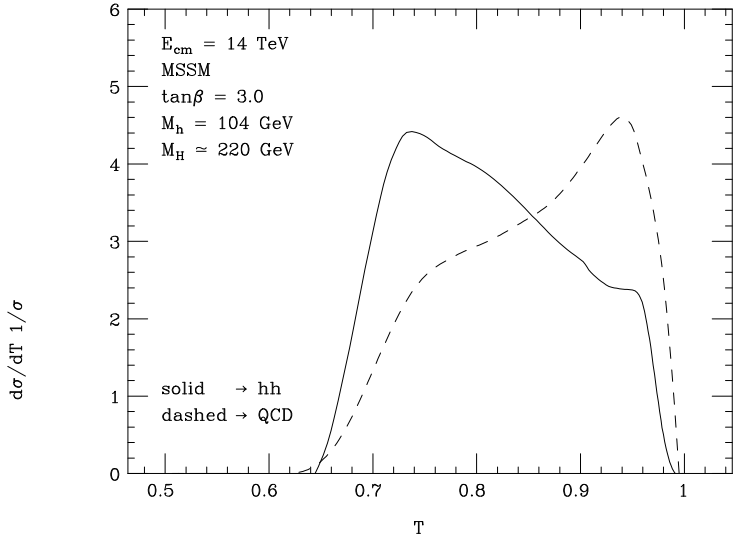}
\caption{
  Distributions in thrust in the rest frame of the $4b$-system
  in $gg\to hh\to b\bar b b\bar b$ and in the QCD background, after
  the cuts (\ref{pTbcut_LHC})--(\ref{Mbbcut_LHC}) at the LHC, for
  $\tan\beta=3$, $m_h=104$ GeV and $m_H\simeq220$ GeV.  Normalisation
  is to unity.}
\label{fig:thrust_LHC}
\end{figure}

An additional consequence that one should expect from the presence
of two intermediate massive objects in $gg\to hh\to b\bar b b\bar b$
events is the spherical appearance of the jets in the final
state, in contrast to the usual planar behaviour of the infrared
QCD interactions. These phenomena can be appreciated in
Fig.~\ref{fig:thrust_LHC}. Notice there the strong tendency
of the background to yield high thrust configurations, again controlled
by the separation cuts when $T$ approaches unity.
On the contrary, the average value of the thrust in the signal
is much lower, being the effect of accidental pairings of `wrong'
$2b$-pairs (the shoulder at high thrust values) marginal. An effective
selection cut seems to be, e.g., $T<0.85$.

\begin{figure}[!ht]
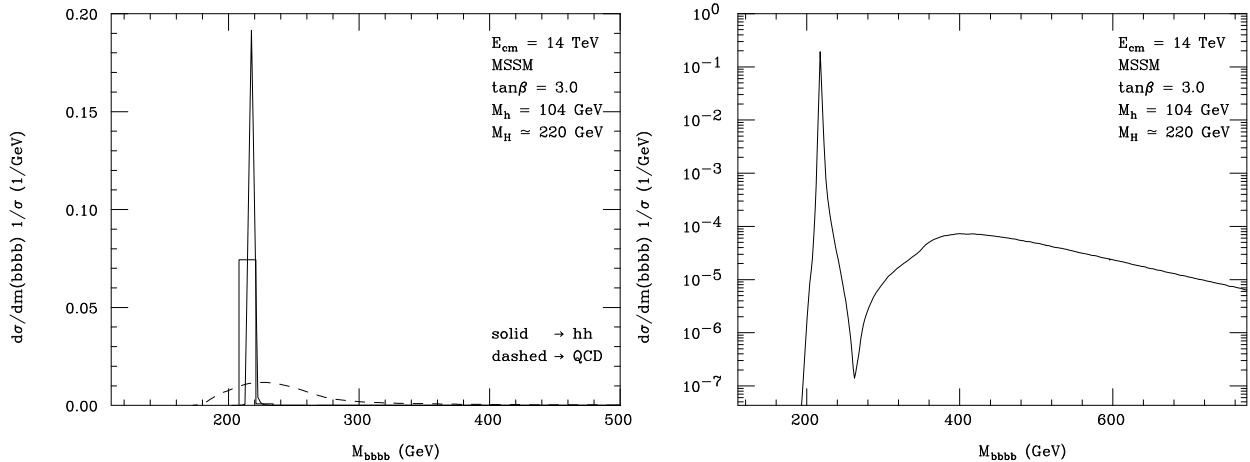

\begin{minipage}[b]{.495\linewidth}
\centering\epsfig{file=Mbbbb_LHC.ps,angle=90,height=6cm,width=\linewidth}
\end{minipage}\hfill\hfill
\begin{minipage}[b]{.495\linewidth}
\centering\epsfig{file=Mbbbblog_LHC.ps,angle=90,height=6cm,width=\linewidth}
\end{minipage}
\caption{ Distributions in invariant mass of the $4b$-system
  in $gg\to hh\to b\bar b b\bar b$ and in the QCD background, after
  the cuts (\ref{pTbcut_LHC})--(\ref{Mbbcut_LHC}) at the LHC, for
  $\tan\beta=3$, $m_h=104$ GeV and $m_H\simeq220$ GeV.  Normalisation
  is to unity. The left hand plot shows both the signal (solid curve)
  and the QCD background (dashed curve), distributed in 5 GeV bins.
  The same signal is also shown as a histogram for a more
  experimentally realistic binning of 13 GeV. The right hand plot also
  shows the signal (collected in 5 GeV wide bins) on a logarithmic
  scale. Here the structure of the continuum contribution (and its
  destructive interference with the heavy Higgs decay contribution)
  can be seen.}
\label{fig:Mbbbb_LHC}
\vspace*{-3mm}
\end{figure}

Furthermore, if the heavy Higgs mass is sufficiently well measured at
the LHC then one can exploit the large fraction \cite{pp} of
$4b$-events which peak at $m_H$ in the signal, as dictated by the
$H\to hh$ decay, improving the signal-to-background ratio.  This peak
at $m_H$ can be clearly seen in the left hand plot of
Fig.~\ref{fig:Mbbbb_LHC}, where it dominates the QCD background, even
for bins 13 GeV wide.  In fact, not only could the QCD background be
considerably suppressed but also those contributions to $gg\to hh$ not
proceeding through an intermediate $H$ state should be removed, this
greatly enhancing the sensitivity of the signal process to the
$\lambda_{Hhh}$ coupling. This can be seen in the right hand plot of
Fig.~\ref{fig:Mbbbb_LHC} where the signal is shown on a logarithmic
scale. The continuum contribution due to the box graphs (and its
destructive interference with the heavy Higgs decay contribution) is
now evident although one should note that it is considerably suppressed
compared to the peak at $m_H$.

Now, if a less than 10\% mass resolution can be achieved on the light
and heavy Higgs masses, then one can tighten cut~(\ref{Mbbcut_LHC}) to
$|m({bb})-m_h|<10$ GeV and introduce the additional cut
$|m({bbbb})-m_H|<20$ GeV. These cuts taken together with those in
$\theta({bb})$ and $T$ already suggested, reduce the QCD background to
the same level as the signal. In fact, we have found that the cross
section of the background drops to approximately 174 fb whereas that
of the signal remains as large as 126 fb, this yielding a very high
statistical significance at high luminosity. Even for less optimistic
mass resolutions the signal-to-background ratio is still significantly
large. For example, selecting events with $|m(bb)-m_h|<20$ GeV and
$|m(bbbb)-m_H|<40$ GeV, the corresponding numbers are approximately
102 fb for the signal and 453 fb for the background. Notice that the
signal actually decreases as these Higgs mass windows are made larger.
This is due to our insistence that exactly two $b \bar b$ pairs should
reconstruct the light Higgs mass. As the light Higgs mass window is
enlarged from $m_h \pm 10$~GeV to $m_h \pm 20$~GeV, it becomes more
likely that accidental pairings reconstruct the light Higgs boson.
Since one is then unable to unambiguously assign the $b$ quarks to the
light Higgs bosons, the event is rejected and the signal drops.

Although we have discussed here an ideal situation which is difficult
to match with more sophisticated hadronic and detector simulations, it
still demonstrates that the measurement of the $\lambda_{Hhh}$
coupling could be well within the potential of the LHC, at least for
our particular choice of MSSM parameters.  Comforted by such a
conclusion, we now move on to more realistic studies.


\subsubsection{$gg\to hh\to b\bar{b}b\bar{b}$ at the LHC experiments}
\label{subsubsec_hadron_LHC}

Although the LHC experiments will be the first where one can attempt
to measure the Higgs self-couplings, the analysis is very challenging
because of the smallness of the production cross sections.  Even in
the most favourable cases, the production rate is never larger than a
few picobarns, already including one-loop QCD corrections, as computed
in Ref.~\cite{spira}. The cross sections at this accuracy are given in
Tab.~\ref{tab:cross}, for the resonant process (case 1 with $m_H = 220$ GeV) 
as well as
three non resonant scenarios: one at the same $\tan\beta$ but with
the $H\to hh$ decay channel closed (case 2), a second at very large
$\tan\beta$ and no visible resonance (case 3) and, finally, the SM
option (case 4, where $m_h$ identifies with the mass of the standard
Higgs state).

\begin{table}[!ht] 
\begin{center}
\begin{tabular}{|l||c|c|c|c|c|c|c|c|c|} \hline
case & model & $\tan\beta$ & $m_h$ (GeV) 
& $A$ (TeV) & 
$\mu$ (TeV) & $\sigma$ (fb) & dominant mode \\ \hline
1  & MSSM  & 3       & 104   
& $+1$   & $-1$   & 2000  & $gg \to H \to hh$ \\
2  & MSSM  & 3       & 100   
& $+1$   & $-1$   & 20  & $gg \to hh$ \\
3  & MSSM  & 50      & 105   
& $+1$   & $+1$   & 5000 & $gg \to hh$ \\
4  & SM    & -       & 105   
& -   & -   & 40 & $gg \to hh$ \\ \hline
\end{tabular}
\caption{Cross sections for double Higgs production $hh$ 
at the LHC via 
gluon-gluon fusion at NLO accuracy, for three possible configurations
of the MSSM and in the SM as well.
}
\label{tab:cross}
\vspace*{-7mm}
\end{center}
\end{table} 

\subsubsection{LHC trigger acceptance}
\label{subsubsect_trigger}

For $4b$-final states, possible LHC triggers are high $p_T$
electron/muons and jets. As an example, the foreseen ATLAS level 1 trigger
thresholds on $p_T$ and acceptance for a
$4b$-selection (with the four $b$-jets reconstructed in the detector)
are given in Tab.~\ref{tab:btag}, assuming the LHC
to be running at high luminosity.

\begin{table}[!ht]
\begin{center}
\begin{tabular}{|l||c|c|c|c|c|c|c|} \hline
trigger type: & 1 $e$ &  1 $\mu$  & 2 $\mu$ & 1 jet & 3 jets & 4 jets & total\\
$p_T$ in GeV & 30 & 20 & 10 & 290 & 130 & 90 & \\  \hline
case 1, $\epsilon(bbbb)$ in \% & 0.01 & 0.01 & 0.4 & 0.08 & 0.08 & 0.05 & 
0.53 \\ \hline
case 2 & $<0.01$ & $<0.01$ & 2.1 & 2.9 & 3.8 & 4.2 & 8.8\\ \hline
case 3 & $<0.01$ & $<0.01$ & 2.2 & 2.7 & 3.8 & 4.1 & 8.7\\ \hline
case 4 & $<0.01$ & $<0.01$ & 2.0 & 2.5 & 3.3 & 3.6 & 7.8\\ \hline
\end{tabular}
\caption{Kinematical
acceptance of the ATLAS detector to trigger four
$b$-jets 
 (including  detector acceptance) at high luminosity.
}
\label{tab:btag}
\end{center}
\end{table}

\begin{figure}
\vspace*{-9mm}
\begin{center}
  \includegraphics[width=6cm]{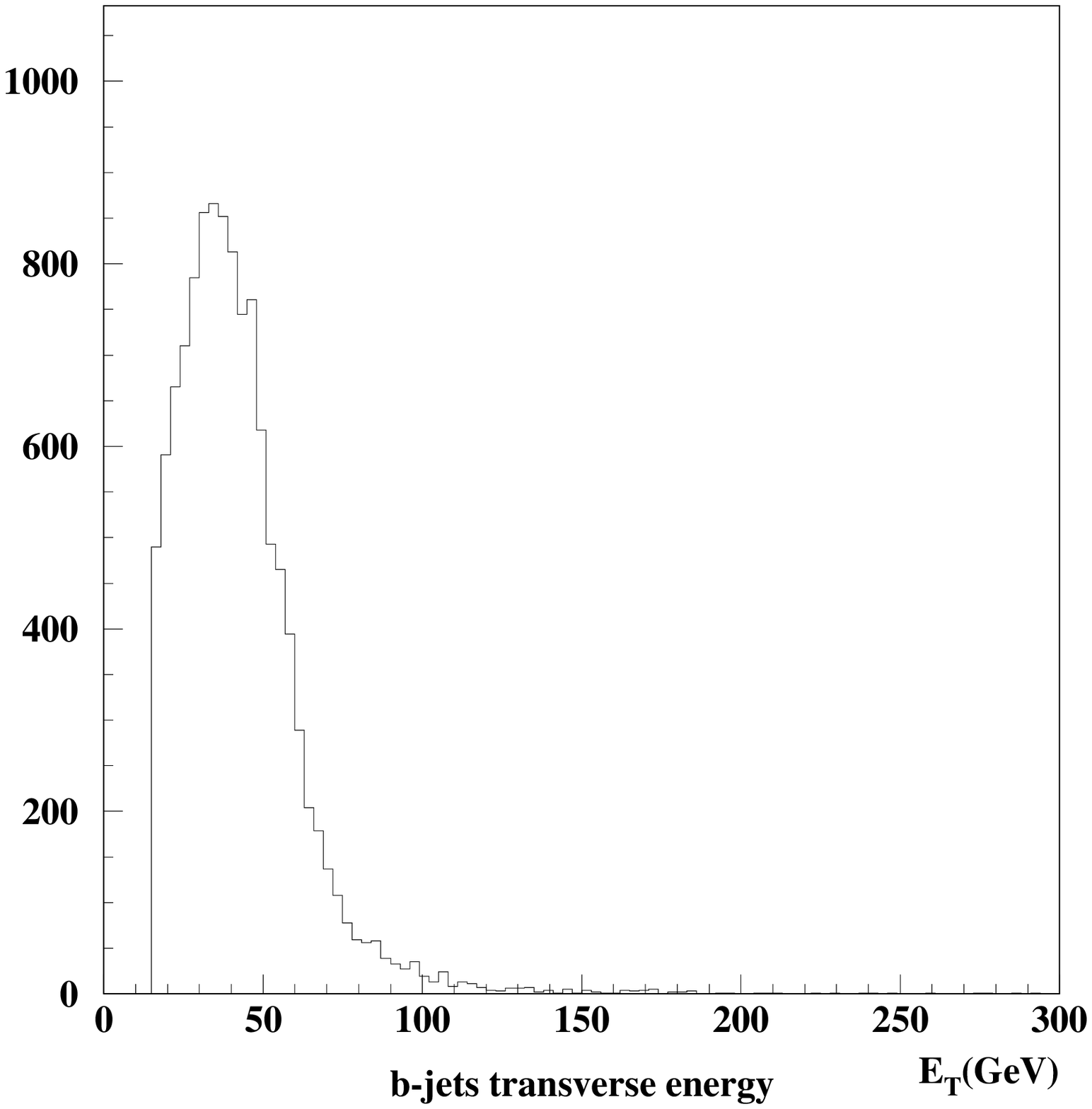}
  \includegraphics[width=6cm]{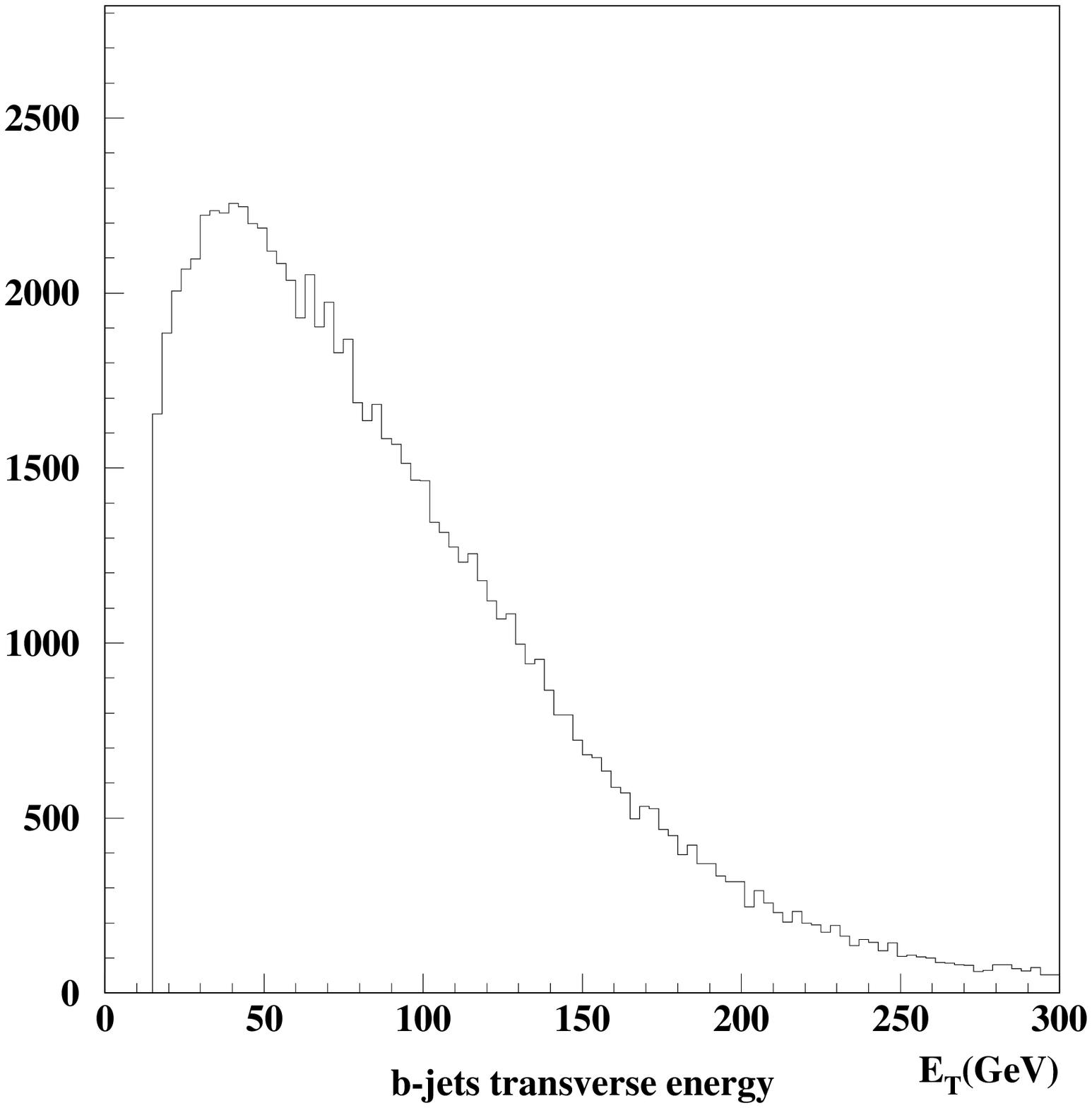}
 \caption{Reconstructed transverse energy/momentum for    
   $b$-jets in $gg\to hh\to b\bar{b}b\bar{b}$ events of case 1
   (left plot) and $b$-jets in $gg\to hh\to b\bar{b}b\bar{b}$ events
   of case 2 (right plot) with ATLAS fast simulation \cite{ATLFAST} at high
   luminosity. Normalisation is arbitrary.}
  \label{ATLF-ptjet-2}
\end{center}
\vspace*{-5mm}
\end{figure}

\noindent
The overall trigger acceptance is at best 8--9\%,
 for cases 2,3,4. The very low efficiency for case 1 is
clearly a consequence of the small value of the difference $m_H-2m_h$,
translating into a softer $p_T(b)$ spectrum with respect to the other
cases (compare the left-hand with the right-hand
side of Fig.~\ref{ATLF-ptjet-2}).
One can further see in the left-hand plot of Fig.~\ref{ATLF-ptjet-2} that 
the bulk of the signal lies below the lowest $p_T(b)$ threshold
of Tab.~\ref{tab:btag} (i.e., $90$ GeV), so that adopting smaller 
trigger thresholds could result in a
dramatic enhancement of our efficiency. Of course, this 
would also substantially increase the low transverse momentum QCD
background, as we can see in the parton level analysis of
Fig.~\ref{fig:pTaftercuts_LHC}. 

For example, by lowering the thresholds to 180, 80
and 50~GeV for 1, 3 and 4 jets, respectively (compare
to Tab.~\ref{tab:btag}), the overall
trigger acceptance on the signal goes up to 1.8\%, i.e., by almost
a factor of 4. Meanwhile though,
the ATLAS level-1 jet trigger rates increase by a factor of 10
\cite{trig}.
Anyhow, even for our poor default value of $\epsilon(bbbb)$ in
Tab.~\ref{tab:btag}, we will see that case 1
still yields a reasonable number of events in the end.
Optimisations of the $b$-jet transverse momentum thresholds
are in progress \cite{more}.

\subsubsection{LHC events selection for $gg\to hh\to b\bar{b}b\bar{b}$}

Jets are reconstructed merging
tracks inside $\Delta R(bb)=0.4$. Only jets with transverse
energy/momentum greater than 30~GeV and with $|\eta(b)|<2.5$ are kept.
(Thus, the same cuts as in the parton level analysis, now applied
instead to jets.)  The effect from pile up is included in the
resolution.  A jet energy correction is then applied.

\begin{figure}[!ht]
\vspace*{-9mm}
\begin{center}
  \includegraphics[width=9cm]{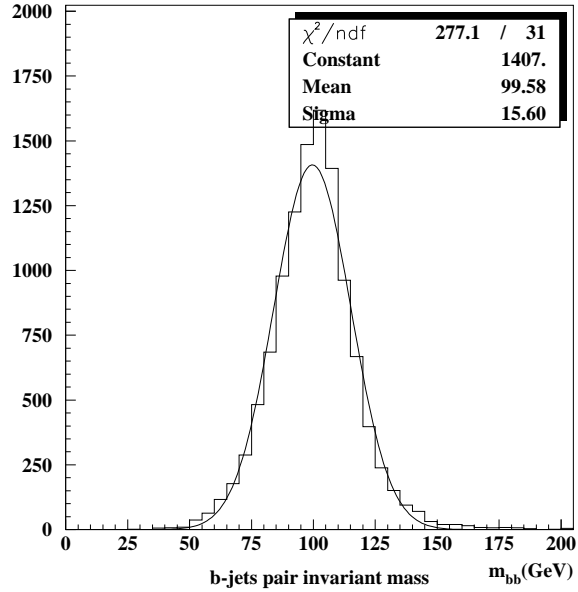} 
  \caption{ Reconstructed invariant mass distribution of
$2b$-jet pairs in continuum $gg\to hh\to b\bar{b}b\bar{b}$ events (case 2) with 
the fast simulation at high luminosity.
Normalisation is arbitrary. (Results of a Gaussian fit are
also given.) }
  \label{ATLF-bbmass-2}
\end{center}
\vspace*{-9mm}
\end{figure}

\begin{figure}[!ht]
\vspace*{-9mm}
\begin{center}
  \includegraphics[width=9cm]{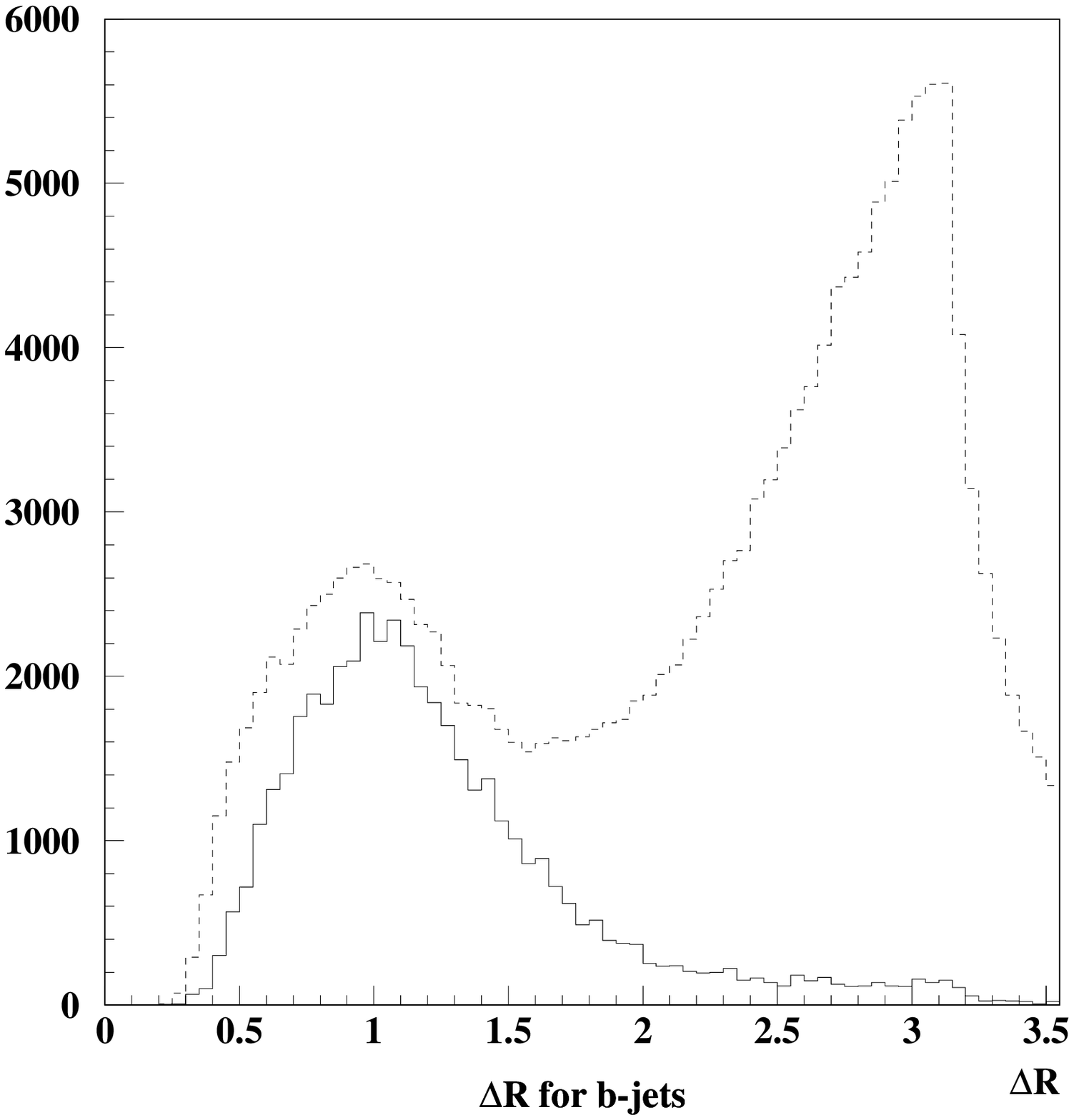} 
  \caption{Reconstructed $\Delta R(bb,bb)$ between
$2b$-jet systems from $h\to bb$ decays 
in continuum $gg\to hh\to b\bar b b\bar b$ events (case 2) with the fast
simulation at high luminosity. The dashed histogram shows the same 
distribution for all pairs of jets.  
Normalisation is arbitrary.}
  \label{ATLF-bbDR-2}
\end{center}
\vspace*{-5mm}
\end{figure}

The invariant masses of each jet pair can then be computed.  Assuming
that the lightest Higgs boson mass is known, events with  $m(bb)$ sufficiently 
close to $m_h$ can efficiently be selected, see
Fig.~\ref{ATLF-bbmass-2}.  Another cut on the $\Delta R(bb,bb)$
between pairs of $b$-jets can also be applied to reduce the intrinsic
combinatorial background, since the latter concentrates at large
$\Delta R(bb,bb)$ values, see Fig.~\ref{ATLF-bbDR-2}.

\begin{figure}[!ht]
\vspace*{-9mm}
\begin{center}
  \includegraphics[width=10cm]{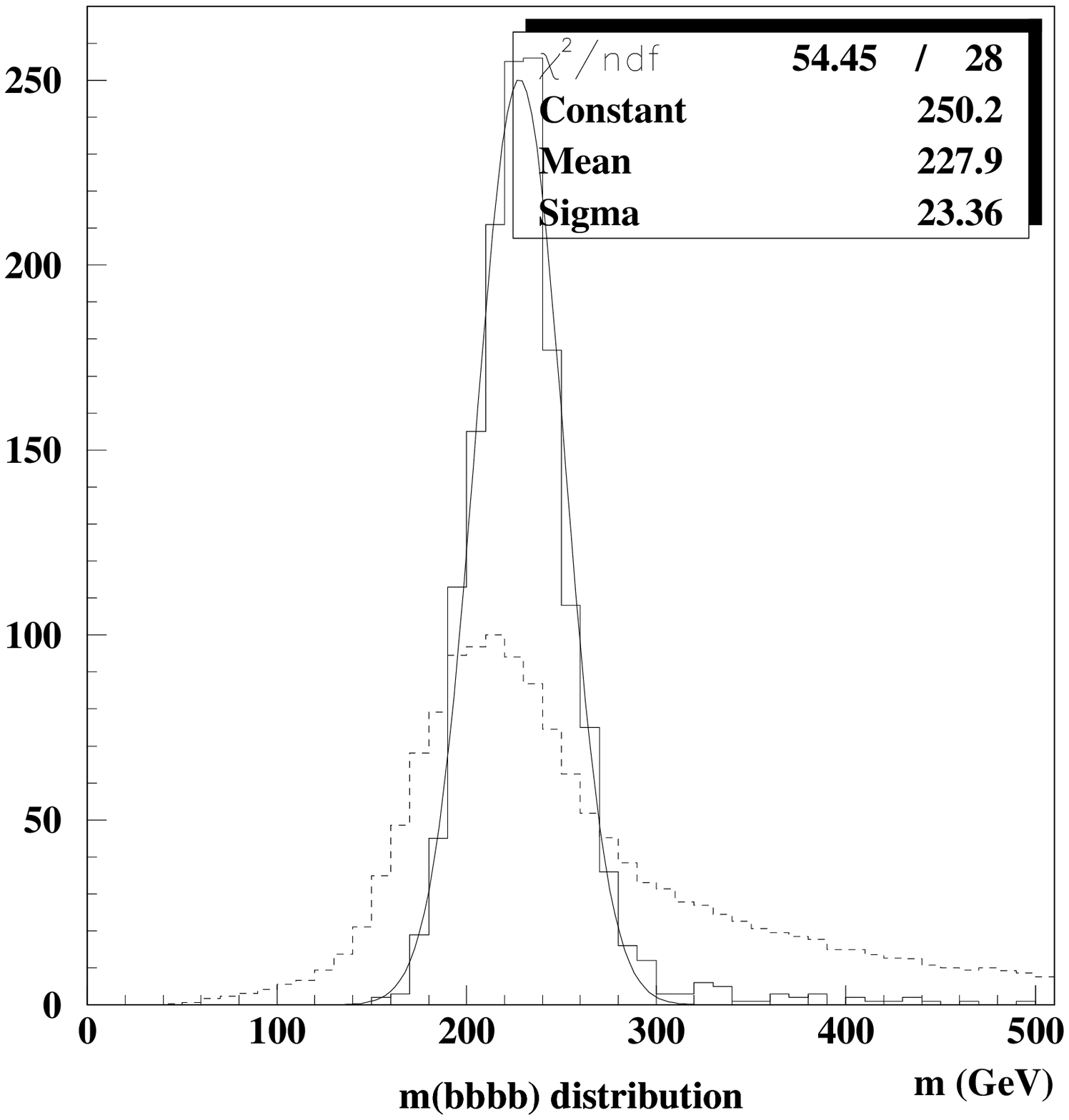} 
  \caption{Reconstructed $4b$-jet invariant mass for
    $b$-jets coming from the $hh$ pair in $gg\to hh\to b\bar bb\bar b$
    events (case 1) with the fast simulation at high luminosity.
    The dashed histogram shows the same distribution for all groups of
    four jets.  Normalisation is arbitrary. (Results of a
    Gaussian fit to the first spectrum are also given.)}
  \label{ATLF-bbbbmass-r-4}
\end{center}
\vspace*{-5mm}
\end{figure}

For case 1, as already discussed, we can further impose that the
invariant mass of the four $b$-jets should be the heavy Higgs mass,
$m_H$, in order to select the $H\to hh$ resonance, as confirmed by
Fig.~\ref{ATLF-bbbbmass-r-4}.  In the other three cases, where the
$H\to hh$ splitting is no longer dominant (MSSM) or non-existent (SM),
one can still insist that the $4b$-jet invariant mass should be higher
than two times the lightest Higgs mass, see 
Fig.~\ref{ATLF-bbbbmass-4} and recall
eq.~(\ref{Mbbbbcut_LHC}).  Finally, following
Fig.~\ref{ATLF-bbbbmass-c-4}, by constraining the $b$-jets pairs
four-momenta around the known light Higgs mass value, $m_h$, one can
further reject the intrinsic background by means of the $m(bbbb)$ spectrum.

\begin{figure}[!ht]
\vspace*{-9mm}
\begin{center}
  \includegraphics[width=10cm]{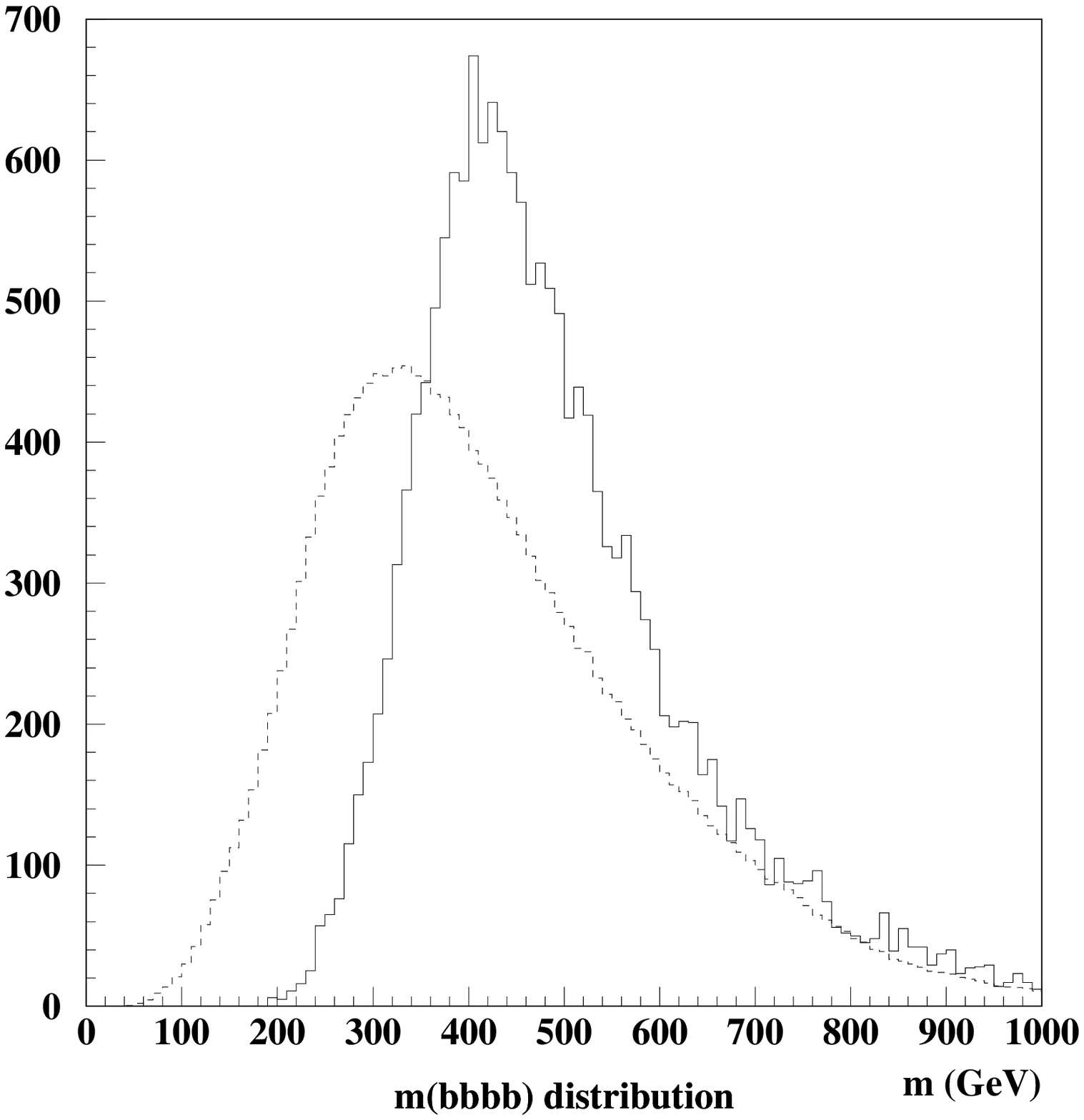} 
  \caption{Reconstructed $4b$-jet invariant mass for
$b$-jets coming from the $hh$ pair in $gg\to hh\to b\bar bb\bar b$ events
(case 4) with the fast simulation at high luminosity.
The dashed histogram shows the same distribution for all groups of four jets.
Normalisation is arbitrary.}
  \label{ATLF-bbbbmass-4}
\end{center}
\vspace*{-5mm}
\end{figure}

\begin{figure}[!ht]
\vspace*{-9mm}
\begin{center}
  \includegraphics[width=10cm]{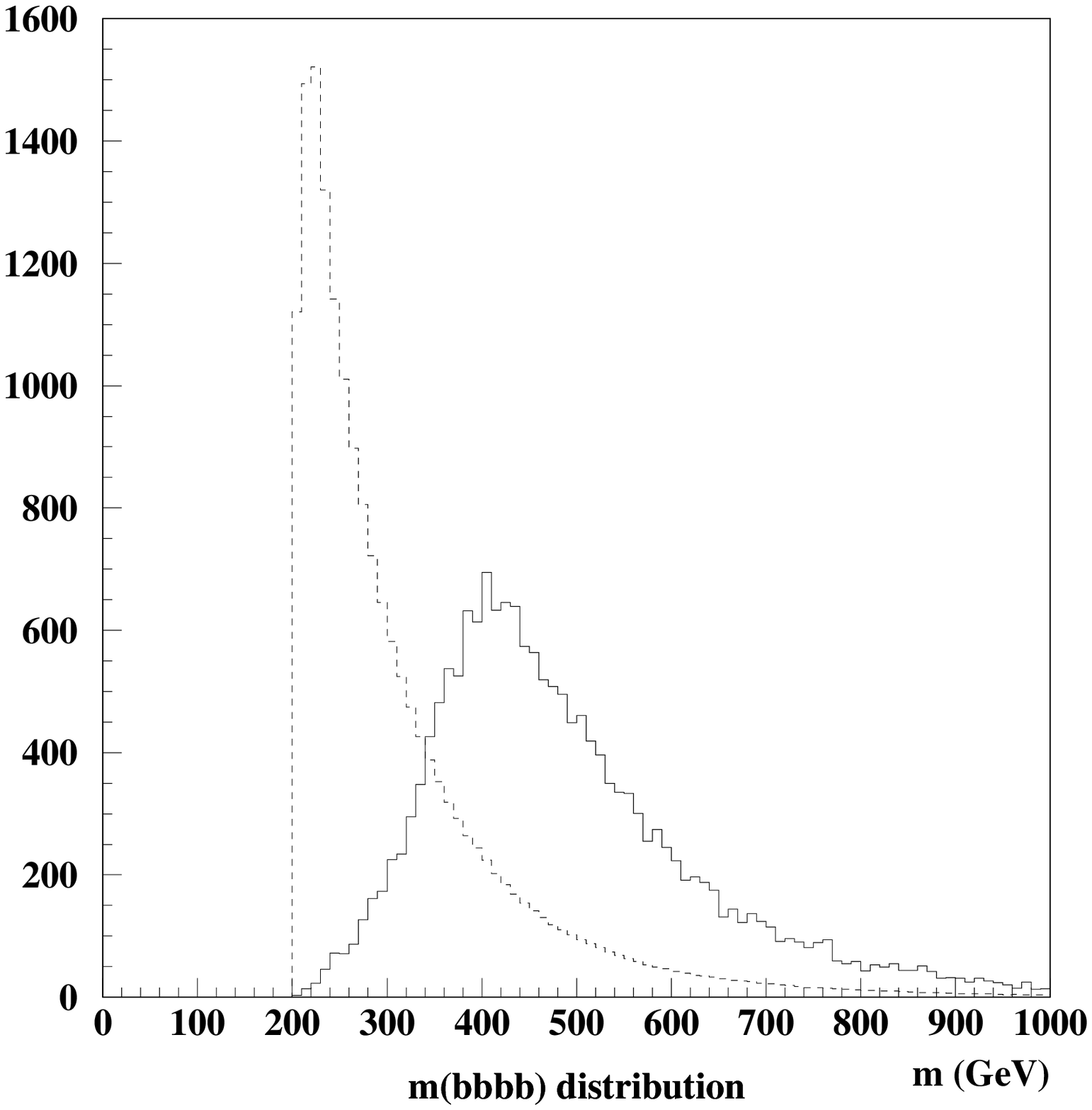} 
  \caption{Reconstructed $4b$-jet invariant mass for
$b$-jets coming from the $hh$ pair in $gg\to hh\to b\bar bb\bar b$ events 
(case 4) with the fast simulation at high luminosity. Here, the 
energy of the jet pairs is recalculated using the $m_h$ constraint.
The dashed histogram shows the same distribution for all groups of four jets.
Normalisation is arbitrary.}
  \label{ATLF-bbbbmass-c-4}
\vspace*{-5mm}
\end{center}
\end{figure}

\subsubsection{LHC $b$-tagging in $gg\to hh\to b\bar{b}b\bar{b}$}

The $b$-tagging efficiency at high luminosity is set to 50\%,
with $p_T$ dependent correction factors for jets rejection. An average
rejection of 10 for $c$-jets and 100 for light-jets can be expected. We then
studied the effect on the selection efficiency of requiring from one
to four $b$-tags, although it is clear that, according to the parton
level studies, the huge background rate demands four $b$-tags, leading
to a 6\% tagging efficiency overall.

\subsubsection{Event rates at the LHC}

Taking into account all the efficiencies described above, and using
the NLO normalisation of Tab.~\ref{tab:cross}, one can extract the
number of expected events per year at the LHC at high luminosity given
in Tab.~\ref{tab:rates}.  The selection cuts enforced here are the
following. For a start, we have kept configurations
where $|m(bb)-m_h|<30$ GeV (cases 1,3,4) or
$|m(bb)-m_h|<20$ GeV (case 2) and $\Delta R(bb,bb)<2.5$ (all four cases).
(If more than two $m_h$'s are reconstructed, the best two $2b$-pairs 
are selected according to the minimum value
of $\delta M^2=[m_h-m(bb)]^2+[m_h-m'(bb)^2]$.) Then, a cut on $m(bbbb)$
is applied: in presence of the $H\to hh$ resonance (case 1) we have
kept events within an $m_H$ mass window of $\pm 2\sigma$ (about 82\%
of the total number survive); otherwise
(cases 2,3,4) we have adjusted the $m(bbbb)\OOrd 2m_h$
cut so to keep 90\% of the sample. In the end, one finds
the numbers in Tab.~\ref{tab:rates}, that are encouraging indeed.

\begin{table}[!ht]
\vspace*{-3mm}
\begin{center}
\begin{tabular}{|l||c|c|c|c|} \hline
                             & case 1 & 2     & 3    & 4 \\ \hline
$\sigma$ in fb                    & 2000   & 20    & 5000 & 40 \\
trigger threshold acceptance       &  0.53\%   & 8.8\% & 8.7\%  & 7.8\% \\
mass windows & 60\%   & 50\%  & 40\% & 40\% \\
$4b$-tagging                  & {6\%}  & {6\%} & {6\%}& {6\%} \\ \hline
events/year (no tagging)     &  636  &  88   & 17400  & 125 \\
events/year (four $b$-tags)  &  38   & 5.3   & 1044   & 7.5 \\ \hline
\end{tabular}
\caption{Total rates for $gg\to hh\to b\bar bb\bar b$, 
after all efficiencies have been included and selection
cuts (\ref{pTbcut_LHC})--(\ref{Rbbcut_LHC}) enforced at hadron level,
with 100 fb$^{-1}$ per year of luminosity.}
\label{tab:rates}
\end{center}
\vspace*{-5mm}
\end{table}

In conclusion then, looking at the results in Tab.~\ref{tab:rates}
and bearing in mind the potential seen in reducing 
 the pure QCD background via $gg\to {\cal O}(\alpha_s^4)\to
b\bar b b\bar b$ (see
Figs.~\ref{fig:thetabb_LHC}--\ref{fig:Mbbbb_LHC}), one should be
confident in the LHC having the potential to measure the
$\lambda_{Hhh}$ coupling in resonant $H\to hh$ events (case 1). To
give more substance to such a claim, we have now initiated background
studies at hadron and detector level, following the guidelines
obtained by the parton level analysis \cite{more}.  As for other
configurations of the MSSM (such as case 2) or in the SM (case
4), the expectations are
more pessimistic. Case 3 deserves
further attention.  In fact, notice the large number of events
surviving and recall what mentioned in the Introduction concerning
 the potential of the non-resonant
$gg\to hh\to b\bar b b\bar b$ process as a discovery channel of the
light Higgs boson of the MSSM in the large $\tan\beta$ region at
moderate $m_A$ values, a corner of the parameter space where the $h$
coverage is given only by SM-like production/decay modes, thus not
allowing one to access information on the MSSM parameters. 
Results on this topic too will be presented in Ref.~\cite{more}.

\subsection{The LC analysis}
\label{subsec_LC}

Here, we closely follow the selection procedure advocated in
Ref.~\cite{noi}.  In order to resolve the four $b$-jets as four
separate systems inside the LC detector region, we impose the
following cuts. First, that the energy of each $b$-jet is above a
minimum threshold,
\begin{equation}\label{Ebcut_LC}
E(b)>10~{\rm{GeV}}.
\end{equation}
Second, that any $b$-jet is isolated from all others, by
requiring a minimum angular separation,
\begin{equation}\label{cosbbcut_LC}
\cos\theta({b,b})<0.95.
\end{equation}
Similarly to the hadronic analysis, one can optimise 
$S/B$ by imposing the constraints
\cite{noi},
\begin{equation}\label{Mbbbbcut_LC}
m({bbbb})\ge 2m_h-10~{\rm{GeV}},
\end{equation}
\begin{equation}\label{Mbbcut_LC}
|m({bb})-m_h|<5~{\rm{GeV}},
\end{equation}
on exactly two combinations of  $2b$-jets.  Here, note that the mass
resolution adopted for the quark systems is significantly better
than in the LHC case, due to the cleanliness of the $e^+e^-$
environment and the expected performance of the LC detectors in jet
momentum and angle reconstruction \cite{resolution}.  Thus, given such
high mass resolution power from the LC detection apparatus, one may
further discriminate between $h$ and $Z$ mass peaks by requiring that
none of the $2b$-jet pairs falls around $m_Z$,
\begin{equation}\label{MZcut_LC}
|m({bb})-m_Z|>5~{\mathrm{GeV}}.
\end{equation}
Moreover, in the double Higgs-strahlung process $e^+e^-\to hhZ$,
the four $b$-quarks are produced centrally,
whereas this is generally not the case for the background (see
the discussion in Ref.~\cite{noi}). This can be exploited by enforcing
\begin{equation}\label{cosbbbbcut_LC}
|\cos\theta({bb,bbb,bbbb})|<0.75,
\end{equation}
where $\theta({bb,bbb,bbbb})$ are the polar angles of all 
two-, three- and four-jet systems.
\begin{figure}[!ht]
~\hskip2.5cm\epsfig{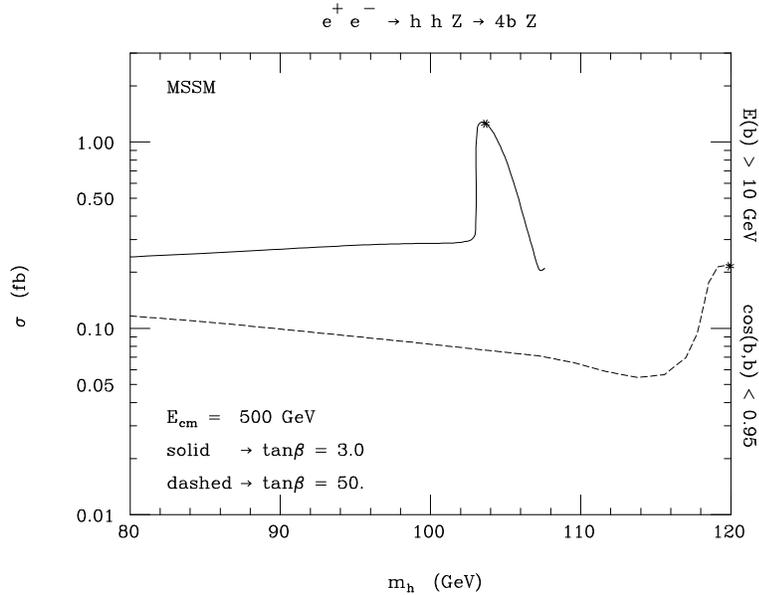}
\caption{Cross sections in femtobarns for the $e^+e^-\to hhZ$ signal
in the $h\to b\bar b b\bar b$ decay channel,
at a LC with 500 GeV as CM energy, as a function of $m_h$
for $\tan\beta=3$ and 50.
Our acceptance cuts in energy and separation of the four $b$-quarks
(\ref{Ebcut_LC})--(\ref{cosbbcut_LC}) have been implemented.
No beam polarisation is included.}
\label{fig:signal_LC}
\end{figure}

Fig.~\ref{fig:signal_LC} shows the production and decay rates of the
signal process, $e^+e^-\to hhZ\to b\bar b b\bar b Z$, as obtained at
the partonic level, after the cuts
(\ref{Ebcut_LC})--(\ref{cosbbcut_LC}) have been implemented. The MSSM
setup here includes some mixing, having adopted $A=2.4$ TeV and
$\mu=1$ TeV, at both $\tan\beta=3$ and 50.  Notice the onset of the
$H\to hh\to b\bar b b\bar b$ decay sequence in the Higgs-strahlung
process $e^+e^-\to HZ$ at low $\tan\beta$.  The same does not occur
for large values, as previously explained. The
impact of the above jet selection cuts on the signal is marginal, as
the $b$-quarks are here naturally isolated and energetic, being the
decay products of heavy objects. In fact, the rates in
Fig.~\ref{fig:signal_LC} would only be 10--20\% higher if all the
$4b$-quark phase space was allowed (the suppression being larger for
smaller Higgs masses). At the height of the resonant peak around
$m_h\approx104$ GeV at $\tan\beta=3$, 
the signal rate is not large but observable,
yielding more than one event every 1 fb$^{-1}$ of data.  For a high
luminosity 500 GeV TESLA design \cite{TESLA}, this would correspond to
more than 300 events per year. Given the very high efficiency expected
in tagging $b$-quark jets, estimated at 90\% for each pairs of heavy
quarks \cite{btag}, one should expect a strong sensitivity to the
triple Higgs self-coupling.  The situation at large $\tan\beta$ is
much more difficult instead, being the production rates smaller by
about a factor of 10.

\begin{figure}[!ht]
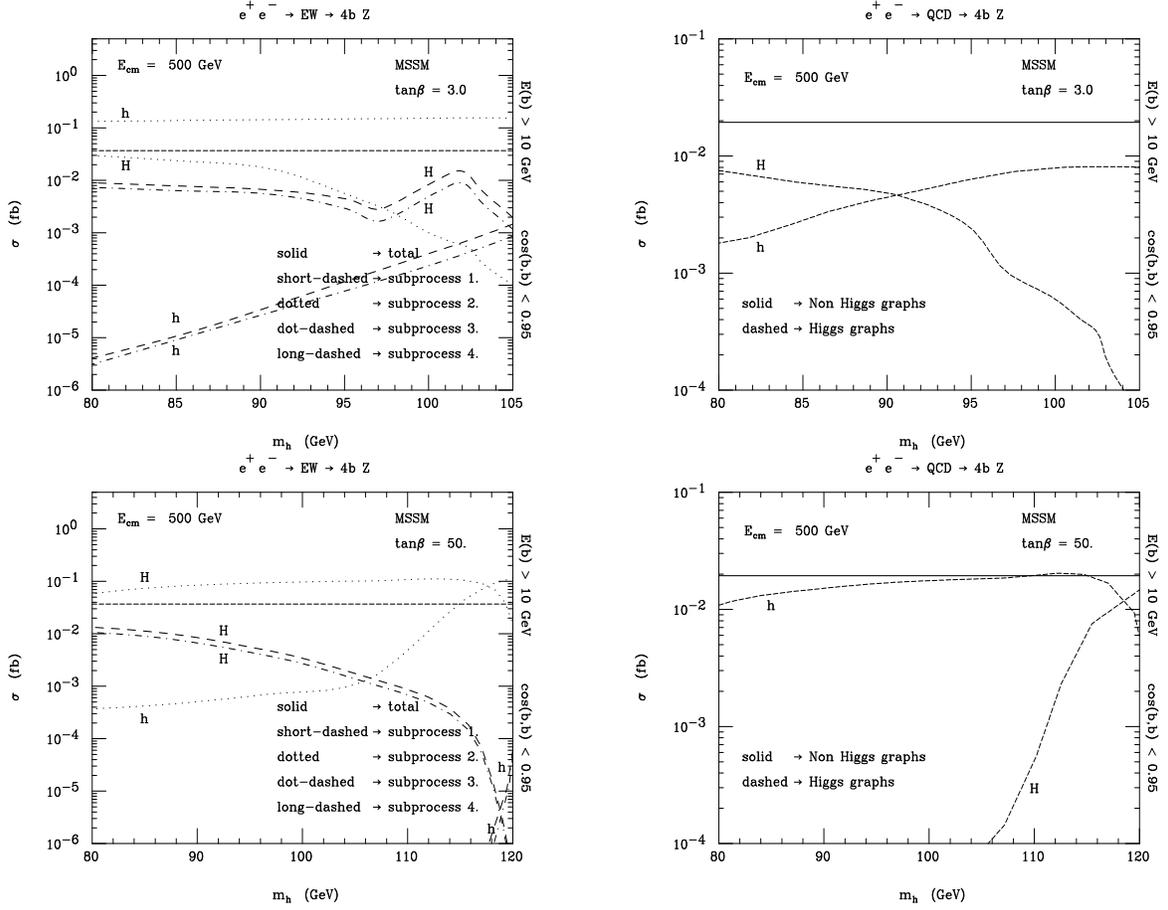

\begin{minipage}[b]{.495\linewidth}
\centering\epsfig{file=split_EW_MSSM_3.ps,angle=90,height=6cm,width=7cm}
\end{minipage}\hfill\hfill
\begin{minipage}[b]{.495\linewidth}
\centering\epsfig{file=split_QCD_MSSM_3.ps,angle=90,height=6cm,width=7cm}
\end{minipage}\hfill\hfill
\begin{minipage}[b]{.495\linewidth}
\centering\epsfig{file=split_EW_MSSM_50.ps,angle=90,height=6cm,width=7cm}
\end{minipage}\hfill\hfill
\begin{minipage}[b]{.495\linewidth}
\centering\epsfig{file=split_QCD_MSSM_50.ps,angle=90,height=6cm,width=7cm}
\end{minipage}
\caption{Cross sections in femtobarns for the dominant components
  of the EW (left) and EW/QCD (right) background to the $e^+e^-\to
  hhZ$ signal in the $h\to b\bar b b\bar b$ decay channel, at a LC
  with 500 GeV as CM energy, as a function of $m_h$ for $\tan\beta=3$
  (top) and 50 (bottom).  Our acceptance cuts in energy and separation
  of the four $b$-quarks (\ref{Ebcut_LC})--(\ref{cosbbcut_LC}) have
  been implemented.  No beam polarisation is included.}
\label{fig:bkgd_LC}
\vspace*{-4mm}
\end{figure}

In the left-hand side of Fig.~\ref{fig:bkgd_LC} we present the EW
background, after the constraints in
(\ref{Ebcut_LC})--(\ref{cosbbcut_LC}) have been enforced, in the form
of the four dominant EW sub-processes.  These four channels are the
following.

\begin{enumerate}
\item $e^+e^-\to ZZZ\to b\bar b b\bar b Z$, first from the left in the
  second row of topologies in Fig.~3 of Ref.~\cite{noi}.  That is,
  triple $Z$ production with no Higgs boson involved.
\item $e^+e^-\to h/HZZ\to b\bar b b\bar b Z$, first(first) from the
  left(right) in the fifth(fourth) row of topologies in Fig.~2 of
  Ref.~\cite{noi} (also including the diagrams in which the on-shell
  $Z$ is connected to the electron-positron line).  That is, single
  Higgs-strahlung production in association with an additional $Z$,
  with the Higgs decaying to $b\bar b$. The cross sections of these
  two channels are obviously identical.
\item $e^+e^-\to h/HZ\to Z^*Z^*Z\to b\bar b b\bar b Z$, first from the
  right in the third row of topologies in Fig.~2 of Ref.~\cite{noi}.
  That is, single Higgs-strahlung production with the Higgs decaying
  to $b\bar b b\bar b$ via two off-shell $Z^*$ bosons.
\item $e^+e^-\to Zh/H\to b\bar b Z^*Z\to b\bar b b\bar b Z$,
  first(first) from the right(left) in the first(second) row of
  topologies in Fig.~2 of Ref.~\cite{noi}.  That is, two single
  Higgs-strahlung production channels with the Higgs decaying to
  $b\bar b Z$ via one off-shell $Z^*$ boson. Also the cross sections
  of these two channels are identical to each other, as in 2.
\end{enumerate}

The ${\cal O}(\alpha_s^2\alpha_{em}^3)$ EW/QCD background is dominated
by $e^+e^-\ar ZZ$ production with one of the two $Z$ bosons decaying
hadronically into four $b$-jets. This subprocess corresponds to the
topology in the middle of the first row of diagrams in Fig.~{4} of
Ref.~\cite{noi}. Notice that Higgs graphs are involved in this process
as well (bottom-right topology in the mentioned figure of \cite{noi}).
These correspond to single Higgs-strahlung production with the Higgs
scalar subsequently decaying into $b\bar bb\bar b$ via an off-shell
gluon. Their contribution is not entirely negligible, owing to the
large $ZH$ production rates, as can be seen in the right-hand side of
Fig.~\ref{fig:bkgd_LC}. The interferences among non-Higgs and Higgs
terms are always negligible.

In performing the signal-to-background analysis, we have chosen two
representative points only, identified by the two following
combinations: (i) $\tan\beta=3$ and $m_A=210$ GeV (yielding
$m_h\approx104$ GeV and $m_H\approx220$ GeV); (ii) $\tan\beta=50$ and
$m_A=130$ GeV (yielding $m_h\approx120$ GeV and $m_H\approx130$ GeV).
These correspond to the two asterisks in Fig.~\ref{fig:signal_LC},
that is, the maxima of the signal cross sections at both $\tan\beta$
values.  The first corresponds to resonant $H\to hh$ production,
whereas the latter to the continuum case.  If we enforce the
constraints of eq.~(\ref{Mbbbbcut_LC})--(\ref{cosbbbbcut_LC}), the
suppression of both EW and EW/QCD is enormous, so that the
corresponding cross sections are of ${\cal O}(10^{-3})$ fb, while the
signal rates only decrease by a factor of four at most.  This is the
same situation that was seen for the SM case in Ref.~\cite{noi}.
Indeed, in the end it is just a matter of how many signal events
survive, the sum of the backgrounds representing no more than a 10\%
correction (see Fig.~11 of Ref.~\cite{noi}). For example, after 500
fb$^{-1}$ of data collected,
one is left with 156 and 15 events for case (i) and (ii),
respectively. However, these numbers do not yet include $b$-tagging
efficiency and $Z$ decay rates.


\section{Summary}
\label{sec_conclusions}

To summarise, the `double Higgs production' subgroup has contributed
to the activity of the Higgs WG by assessing the feasibility of
measurements of triple Higgs self-couplings at future TeV colliders.
The machines considered were the LHC at CERN (14 TeV) and a future LC
running at 500 GeV.  In both cases, a high luminosity setup was
assumed, given the smallness of the double Higgs production cross
sections.  In particular, the $H\to hh$ resonant enhancement was the
main focus of our studies, involving the lightest, $h$, and the
heaviest, $H$, of the neutral Higgs bosons of the MSSM, in the
kinematic regime $m_H\OOrd2m_h$. This dynamics can for example occur
in the following reactions: $gg\to hh$ in the hadronic case and
$e^+e^-\to hhZ$ in the leptonic one, but only at low $\tan\beta$.
These two processes proceed via intermediate stages of the form $gg\to
H$ and $e^+e^-\to HZ$, respectively, followed by the decay $H\to hh$.
Thus, they in principle allow one to determine the strength of the
$Hhh$ vertex involved, $\lambda_{Hhh}$, in turn constraining the form
of the MSSM Higgs potential itself. The signature considered was
$hh\to b\bar b b\bar b$, as the $h\to b\bar b$ decay rate is always
dominant.

We have found that several kinematic cuts can be exploited in order to
enhance the signal-to-background rate to level of high significance,
particularly at the $e^+e^-$ machine. At the $pp$ accelerator, in
fact, the selection of the signal is made much harder by the presence
of an enormous background in $4b$ final states due to pure QCD. In
parton level studies, based on the exact calculation of LO scattering
amplitudes of both signals and backgrounds (without any showering and
hadronisation effects but with detector acceptances), we have found
very encouraging results. At a LC, the double Higgs signal can be
studied in an essentially background free environment. 
 At the LHC, the signal and the QCD background are in the end
at the same level with detectable but not very large cross sections.

Earlier full simulations performed for the $e^+e^-$ case had already
indicated that a more sophisticated treatment of both signal and
backgrounds, including fragmentation/hadronisation and full detector
effects, should not spoil the results seen at the parton level.  For the
LHC, our preliminary studies of $gg\to H\to hh\to b\bar b b\bar b$
in presence of
the $gg\to hh\to b\bar b b\bar b$ continuum (and relative interferences)
also point to the feasibility of the signal selection, after realistic
detector simulation and event reconstruction. 
As for double $h$ production in the continuum, 
although not very useful for Higgs self-coupling measurements,
this seems a promising channel, if not to discover the
lightest MSSM Higgs boson certainly to study its properties 
and those of the Higgs sector in general
 (because of the large production
and decay rates at high $\tan\beta$ and its sensitivity to such a parameter),
as shown from novel simulations also presented in this study. 
(The discovery potential of this mode will eventually be addressed
in Ref.~\cite{more}.)
Despite lacking a full
background analysis in the LHC case, we have no reason to believe that
a comparable degree of suppression of background events seen at parton level
cannot be achieved also at hadron level.  Progress in this respect is
currently being made \cite{more}.

\subsubsection*{Acknowledgements}

SM acknowledges financial support from the UK-PPARC.  The authors
thank Patrick Aurenche and the organisers of the Workshop for the
stimulating environment that they have been able to create. DJM and MM
thank Michael Spira for useful discussions. Finally, we all thank Elzbieta
Richter-Was for many useful comments and suggestions.

\end{document}